\documentclass[twocolumn,showpacs,amsmath,amssymb,amsfonts,floatfix,prl,aps]{revtex4}
\usepackage{graphicx}
\usepackage{color}
\usepackage{braket}
\usepackage{algorithm}
\usepackage{algpseudocode}
\usepackage{framed}

\begin{document}
\title{An efficient algorithm for time propagation within time-dependent density functional theory}
\author{J. K. Dewhurst$^{1}$}
\author{K. Krieger$^{1}$}
\author{S. Sharma$^{1,2}$}
\email{sharma@mpi-halle.mpg.de}
\author{E. K. U. Gross$^{1}$}
\affiliation{1. Max-Planck-Institut f\"ur Mikrostrukturphysik, Weinberg 2, 
D-06120 Halle, Germany.}
\affiliation{2. Department of physics, Indian Institute for technology-Roorkee, 247667 Uttarkhand, India}
\date{\today}

\begin{abstract}
An efficient algorithm for time propagation of the time-dependent Kohn-Sham equations is presented. The algorithm is based on dividing
the Hamiltonian into small time steps and assuming that it is constant over these steps. This allows for the
time-propagating Kohn-Sham wave function to be expanded in the instantaneous eigenstates of the Hamiltonian. The stability and 
efficiency of the algorithm are tested not just for non-magnetic but also for fully non-collinear magnetic systems. 
We show that even for delicate properties, like magnetization density, large time-step sizes can be used indicating
the stability and efficiency of the algorithm. 
\end{abstract}

\maketitle

%%%%%%%%%%%%%%
% Introduction
%%%%%%%%%%%%%%

Manipulation of electrons by ultra-short laser pulses will ultimately lead to ultra-fast devices. 
In order to design such devices without actually performing experiments,
one needs an \emph{ab-inito} theory for treating real materials under the influence of time-dependent external fields.
Time-dependent density functional theory (TDDFT)\cite{runge84},
which extends density functional theory into the time domain, is a formally exact method for describing the real-time 
dynamics of interacting electrons. An essential element in solving a problem using 
TDDFT on a computer is an algorithm to propagate the time dependent Schr\"odinger equation:
\begin{eqnarray}
\label{se}
i\frac{\partial}{\partial t} \left| \Phi_i(t) \right> = \hat{H}(t)\left| \Phi_i(t) \right>,
\end{eqnarray}
where $\hat {H}$ is the Hamiltonian and $\Phi$ the wave function of interacting electrons. By the virtue of the Runge-Gross 
theorem\cite{runge84}, one can obtain the exact time-propagation of the density of this fully interacting system by solving single particle 
time-dependent Kohn-Sham (KS) equations. In our particular case, where the orbitals are Pauli spinors, these are
\begin{eqnarray}
\label{hamil}
&i&\frac{\partial \psi_j({\bf r},t)}{\partial t}=\left[
\frac{1}{2}\left(-i\overrightarrow{\nabla} +\frac{1}{c}\overrightarrow{A}_{\rm ext}(t)\right)^2 +v_{s}({\bf r},t) \right. \\ \nonumber
&+&\left. 
\frac{1}{2c} \overrightarrow{\sigma}\cdot\overrightarrow{B}_{s}({\bf r},t) +
\frac{1}{4c^2} \overrightarrow{\sigma}\cdot (\overrightarrow{\nabla}v_{s}({\bf r},t) \times i\overrightarrow{\nabla})\right]
\psi_j({\bf r},t),
\end{eqnarray}
where $\overrightarrow{A}_{\rm ext}(t)$ is a external vector potential, $\overrightarrow{\sigma}$ are the Pauli matrices
and $\psi_j$ are the KS orbitals. 
The KS effective potential $v_{s}({\bf r},t) = v_{\rm ext}({\bf r},t)+v_{\rm H}({\bf r},t)+v_{\rm xc}({\bf r},t)$ is 
decomposed into the external potential $v_{\rm ext}$, the classical electrostatic Hartree potential $v_{\rm H}$ and 
the exchange-correlation (XC) potential $v_{\rm xc}$.  Similarly the KS magnetic field is written as 
${\bf B}_{s}({\bf r},t)={\bf B}_{\rm ext}(t)+{\bf B}_{\rm xc}({\bf r},t)$ where ${\bf B}_{\rm ext}(t)$ is an external
magnetic field and ${\bf B}_{\rm xc}({\bf r},t)$ is 
the XC magnetic field. The final term of Eq. (\ref{hamil}) is the spin-orbit coupling term. Requirements for any 
accurate\cite{moler03,castro04} time-propagation algorithm are 
(a) stability: the errors do not build up as the system is propagated for longer times, 
(b) efficiency: time propagation is performed by dividing the the total time interval into steps and it is essential
for an efficient algorithm to allow for large time steps and
(c) unitarity: which is required for maintaining the normalization of the wave function at each time-step. 
In the following we outline one such algorithm which satisfies all the above criteria and can be easily implemented 
in existing computer codes. 

%%%%%%%%%%%%%%
% algorithm
%%%%%%%%%%%%%%
\begin{figure}[ht]
\centerline{\includegraphics[width=\columnwidth,angle=-0]{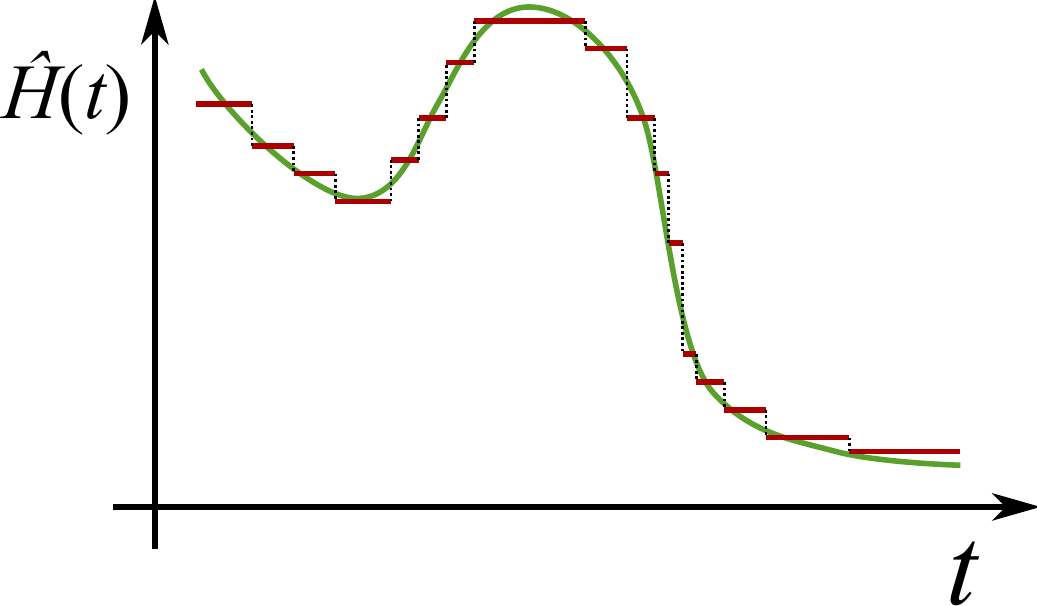}}
\caption{(Color online) Hamitonian as a function of time (full line) and approximation to this hamiltonian (step function).}
\label{ft}
\end{figure}

The solution of the KS equations can be represented by means of the time evolution operator:
\begin{eqnarray}
\ket{\psi_i(T)}=\hat{U}(T,0)\ket{\psi_i(0)},
\end{eqnarray}
where $\hat{U}(T,0)$ is the time evolution operator that propagates all TD-KS states from time $t=0$ to the final time $t=T$.
The time evolution operator satisfies the composition law:
\begin{eqnarray}
\hat{U}(T,0) = \hat{U}(T,T-\Delta t) \dots \hat{U}(2\Delta t,\Delta t) \, \hat{U}(\Delta t,0)
\end{eqnarray}
which allows for division of the total time propagation into small steps of step length $\Delta t$. 
In the limit $\Delta t \rightarrow 0$ this time propagation operator can be written as:
\begin{eqnarray}
 \hat{U}(t+\Delta t,t) = e^{-i\hat{h}_{s}(t)\Delta t}.
\end{eqnarray}
In principle this exponential expression can be used to stepwise propagate all TD-KS states, in practice however, such an
exponential expression of an operator is nearly impossible to calculate exactly (except in certain trivial cases) and iterative
schemes like polynomial expansion\cite{yabana96,talezer84,baer01,chen99}, Krylov subspace projection\cite{park86,hochbruk97} 
and splitting techniques are used\cite{feit82-1,feit82-2,suzuki93,suzuki92,mikhailova99,bandrauk93,sugino99,watanabe02,milfeld83}.
All these techniques have been tried and tested, mainly for finite systems, and each one has its own set of advantages and disadvantages\cite{castro04}.

In the present work we propose a new method for time propagation in which the Hamiltonian is divided into time steps ($\Delta t$) 
and it is assumed that the Hamiltonian remains constant 
between time $t$ and $t+\Delta t$ (see Fig. 1). If this can be done then the time evolution operator in the basis of the instantaneous eigenstates
of $H$ trivially becomes
\begin{eqnarray}
 \hat{U}(t+\Delta t,t) = e^{-i\epsilon(t)\Delta t},
\end{eqnarray}
where $\epsilon(t)\equiv {\rm diag}(\epsilon_1(t), \cdots, \epsilon_n(t))$ are the instantaneous eigenvalues. 
Thus if the Hamiltonian can be diagonalized at each time step, the time propagating KS states can be expanded in 
instantaneous eigenstates of the Hamiltonian. This algorithm is particularly
suited for codes where full diagonalization is performed and can be outlined as follows.
Let $\chi_i({\bf r})$ be the ground state Kohn-Sham orbitals at $t=0$ and set $c_{ij}(t=0)=\delta_{ij}$.

\begin{center}
\begin{framed}
 \begin{algorithmic}[1]
 \State Set $\psi_j({\bf r},t)=\sum_i c_{ij}(t)\chi_i({\bf r})$
 \State Compute $\rho({\bf r},t)$ and ${\bf m}({\bf r},t)$
 \State Compute $v_s({\bf r},t)$, ${\bf B}_s({\bf r},t)$, ${\bf A}_s({\bf r},t)$ to give $\hat{H}(t)$
 \State Compute $H_{ij}\equiv\bra{\chi_i}\hat{H}(t)\ket{\chi_j}$
 \State Solve $H_{ik}d_{kj}=\epsilon_j d_{ij}$ for $d$ and $\epsilon$
 \State Compute $c_{ij}(t+\Delta t) =\sum_{kl} d_{jk}^* d_{lk}\,e^{-i\epsilon_k \Delta t} c_{il}(t)$
 \State If $t<T$ goto step 1
 \end{algorithmic}
\end{framed}
\end{center}
Here $\rho$ is the charge density and ${\bf m}$ is the magnetisation density; and the potentials $v_s$, ${\bf B}_s$ and ${\bf A}_s$
are functionals of these densities.
It is important to mention that this algorithm is unitary and thus the KS orbitals are orthonormal at each time-step. 

%%%%%%%%%%%%%%
% elk details 
%%%%%%%%%%%%%%
For testing the validity of the algorithm outlined above, various extended systems are studied\cite{param}
using the full-potential linearized augmented plane wave (FP-LAPW)
method \cite{Singh} as implemented within the Elk code \cite{elk}. 

%%%%%%%%%%%%%%
% algo test 1
%%%%%%%%%%%%%%
\begin{figure}[ht]
\centerline{\includegraphics[width=\columnwidth,angle=-0]{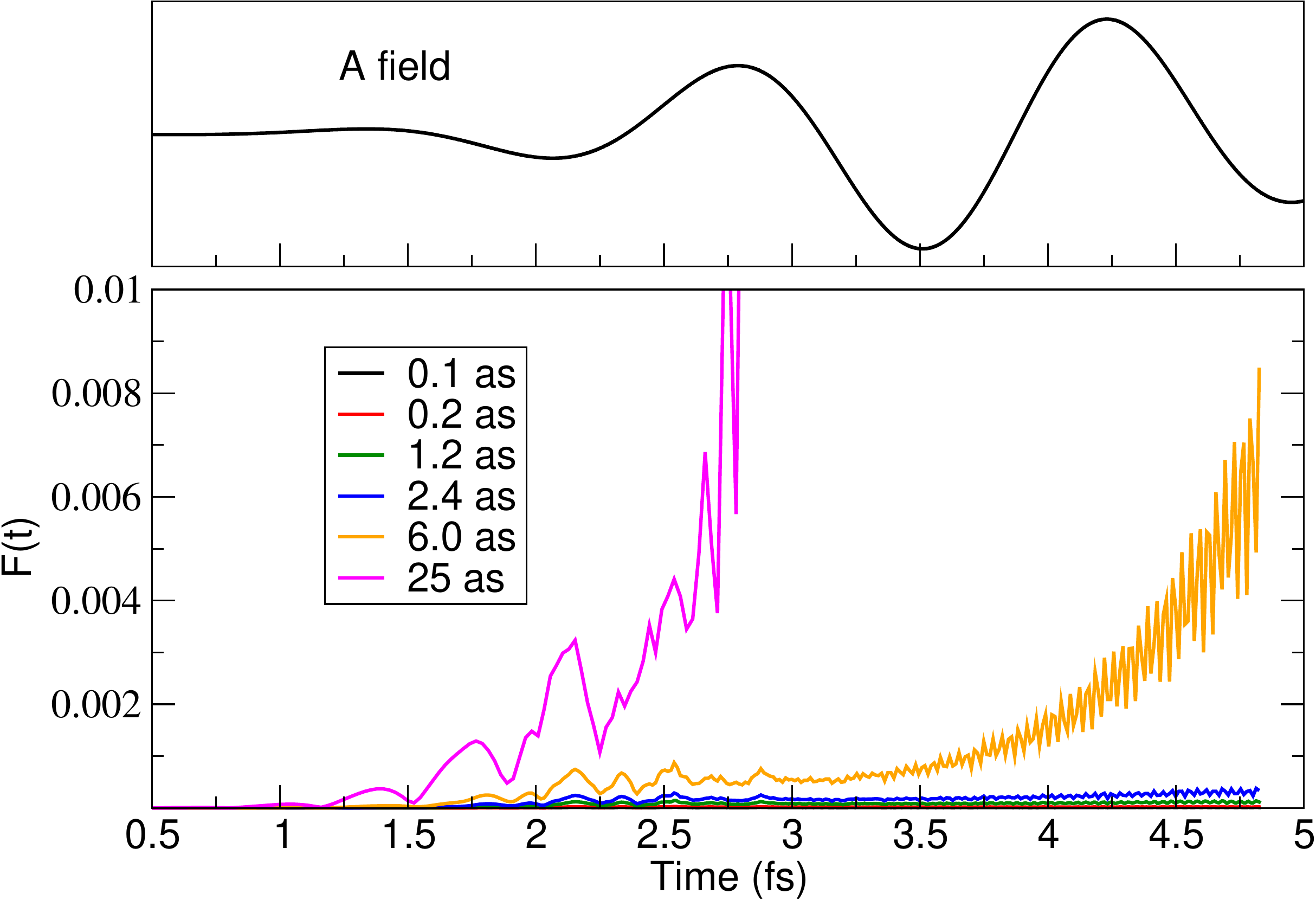}}
\caption{(Color online) Upper panel: vector potential, $A(t)$, of the applied laser field.  
Lower panel: function $F(t)$ (as defined in Eq. \ref{func}), for various time steps, 
as a function of time (in femtoseconds).}
\label{ft}
\end{figure}
\begin{figure}[ht]
\centerline{\includegraphics[width=\columnwidth,angle=-0]{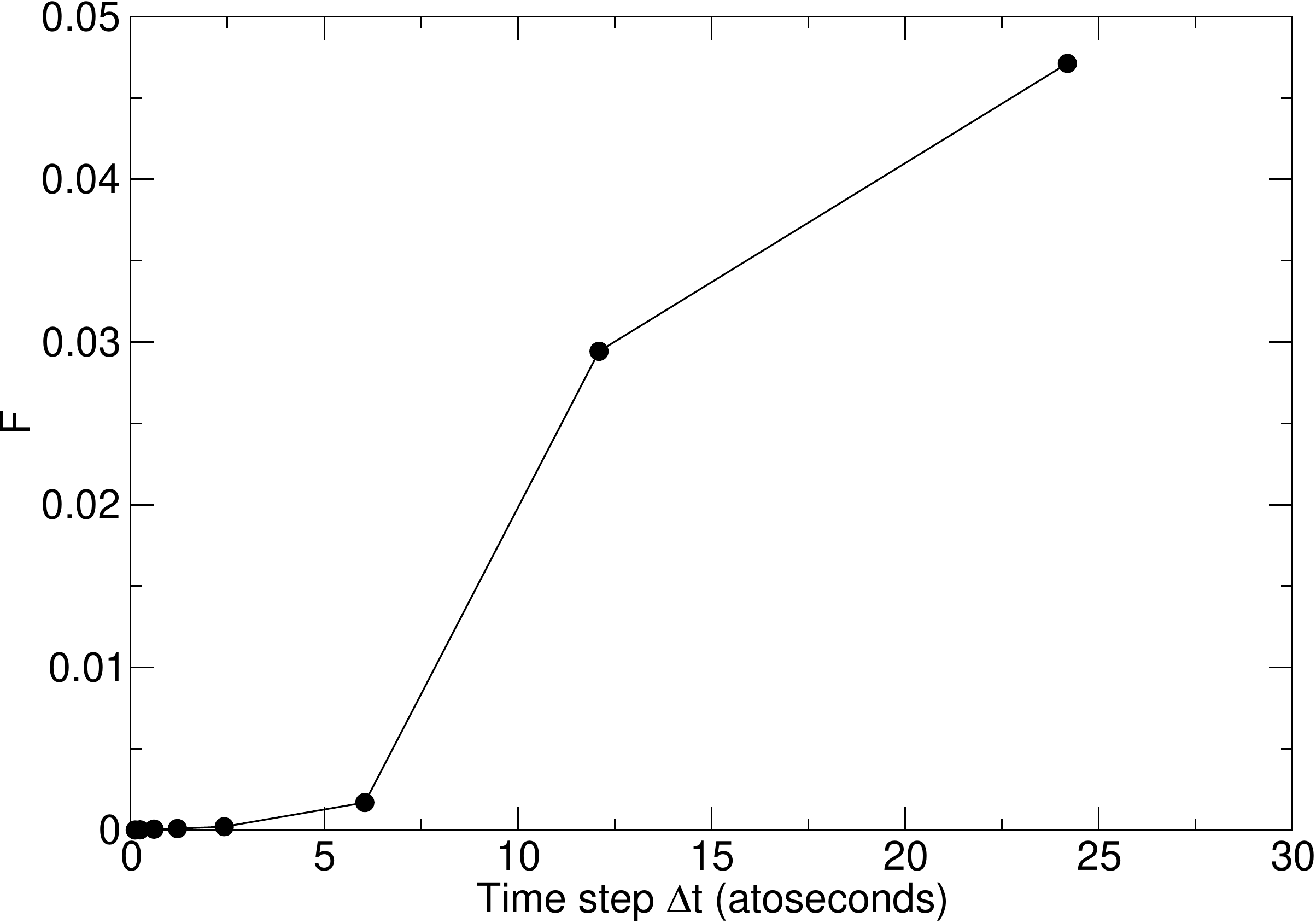}}
\caption{Function $F(t)$ integrated over time as a function of time step size ($\Delta t$)}
\label{f}
\end{figure}

The efficiency of this algorithm depends upon the step length $\Delta t$ as well as how easy it is to diagonalize the
Hamiltonian in step 5. In the limit $\Delta t \rightarrow 0$ 
the algorithm is exact. It still remains to be seen how large the time step can be chosen so that small errors
do not build up as the system is propagated for long times.
In order to test this, we first design quantities which will provide a stringent check for efficiency and stability of the algorithm. 
In the following we present one such quantity,
\begin{eqnarray}\label{func}
F(t)=\frac{1}{2N}\int d^3r \left| \rho_1({\bf r},t)-\rho_2({\bf r},t)\right|,
\end{eqnarray}
where $N$ is the number of electrons and $\rho_1$ and $\rho_2$  are the time-dependent charge densities from two different time
propagations of the same Hamiltonian. The difference between these two time propagations is the length of the time step $\Delta t$. 
In the extreme case where the two densities are so different that they do not overlap at any space point then $F(t)=1$ and 
if the two densities are exactly the same then $F(t)=0$. 
Thus deviation of $F(t)$ from 0 is an indicator of the instability of the algorithm.
In Fig. \ref{ft} are plotted $F(t)$ for solid Fe\cite{param} under the influence of a time-dependent external vector 
potential corresponding to an intense laser pulse\cite{pulse}. 
The smallest step length used for time propagation was 0.06 attoseconds (as) (this determines the $\rho_1$). 
It is clear from these results that the error for step sizes below 5 as are negligible and can easily be used to obtain 
reliable results. The errors also do not build up as the system is time propagated over longer times. For step sizes of 6 as 
or greater, the error is large and builds up as the Hamiltonian is propagated for longer times. 

%%%%%%%%%%%%%%
% algo test 2 
%%%%%%%%%%%%%%
While doing large scale practical calculations, it is difficult to look at quantity like $F(t)$ for each case. It is much 
more convenient to integrate $F(t)$ over time and look at this single number as a function of $\Delta t$. 
This is plotted in Fig. \ref{f}. These results again indicate that time step up to 2.5 as can easily be used. 
It is important to mention that for studying time-dependent phenomena in the few hundred femtoseconds regime, a typical step size 
of $\sim 1$ as is used. Usually such systems are 
studied without taking the magnetization density into account, which is much more sensitive quantity than charge
density itself. Despite this we find that large step sizes ($\sim 2.5$ as) can be used for the time propagation which indicates the
stable nature of this algorithm. Similar tests for LiF, a non-magnetic material, reveal that a step size as large as $\sim 6$ 
as can reliably be used. 

\begin{figure}[ht]
\centerline{\includegraphics[width=\columnwidth,angle=-0]{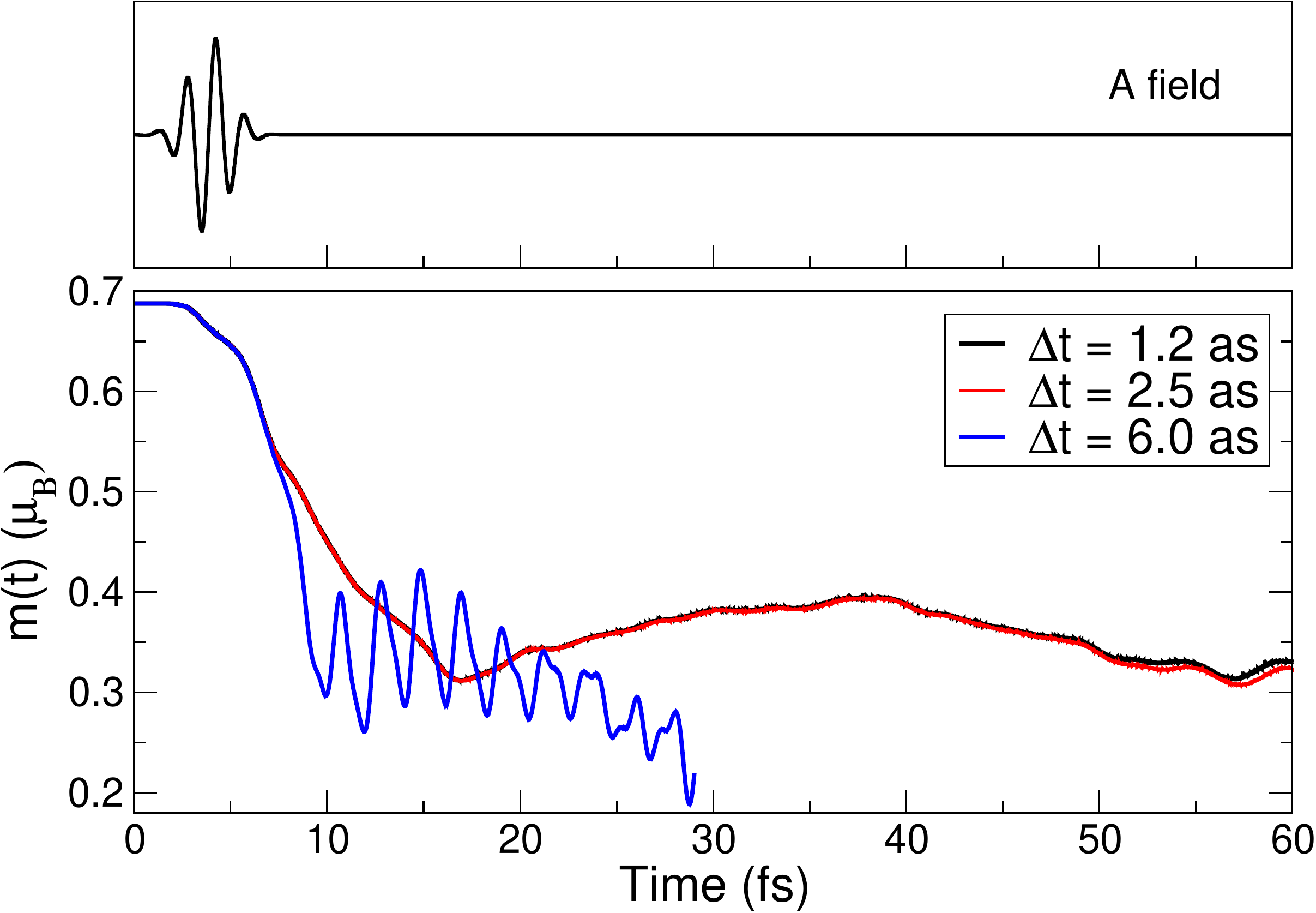}}
\caption{(Color online) Upper panel: vector potential, $A(t)$, of the applied laser pulse.
Lower panel: magentic moment (in Bohr magneton) per Ni atom as a function of time (in femtoseconds) 
for three differet time step sizes.}
\label{mom}
\end{figure}

%%%%%%%%%%%%%%
% algo test 3
%%%%%%%%%%%%%%
One can make the test conditions for the algorithm even more stringent by performing similar tests for a fully non-collinear
system with spin-orbit coupling. Results for
the magnetic moment per atoms for solid Ni\cite{param}  under the influence of external time-dependent vector potential 
of an intense laser pulse\cite{pulse} are shown in 
Fig. \ref{mom}. Here the tests are performed for Ni rather than Fe simply because Ni has delocalized electrons with very 
small moment and is highly sensitive to computational details. 
The system is non-collinear and undergoes demagnetization due to the presence of 
spin-orbit coupling term in Eq. 1. The plotted magnetic moment shows that the step size as large as 2.5 as can 
be used in this case. Not surprisingly, the intensity of the external laser pulse can play an 
important role in determining the step length. The more intense the pulse the smaller the required step length. To give the
algorithm a stringent test, the pulse intensity in the present case is ($10^{15}$ W/cm$^2$) chosen to be the highest 
used for such calculations\cite{our}. It is important to note that in the present work we have
used pulses with wave length in the optical regime. For extreme ultraviolet pulses, 
obviously, the time step $\Delta t$ has to be chosen small enough to resolve the wave length.

%%%%%%%%%%%%%%
% conclusions
%%%%%%%%%%%%%%
To summarize: we present an efficient algorithm for time propagating the Kohn-Sham equations. The algorithm is based on dividing
the full time into small time steps and assuming that the Hamiltonian remains constant over each step. This
allows for expansion of the time-propagating orbitals in the basis of instantaneous eigenstates of the Hamiltonian. 
This algorithm is ideally suited for codes where full diagonalization is performed. By performing stringent tests
for collinear and non-collinear magnetic systems we demonstrate the efficiency of the algorithm.

%\bibliography{algo}
\end{document}